\documentclass[fleqn,usenatbib]{mnras}

\usepackage{newtxtext,newtxmath}

\usepackage[T1]{fontenc}

\DeclareRobustCommand{\VAN}[3]{#2}
\let\VANthebibliography\thebibliography
\def\thebibliography{\DeclareRobustCommand{\VAN}[3]{##3}\VANthebibliography}

\usepackage{graphicx}	
\usepackage{amsmath}	
\usepackage{mathtools}
\usepackage{orcidlink}

\newcommand{\fesc}{\ifmmode{f_{\rm esc}}\else{$f_{\rm esc}$}\fi}
\newcommand{\fescs}{\ifmmode{f_{\rm esc}^\star}\else{$f_{\rm esc}^\star$}\fi}
\newcommand{\kms}{\ifmmode{{\;\rm km~s^{-1}}}\else{km~s$^{-1}$}\fi}
\newcommand{\fgas}{\ifmmode{{f_{\rm gas}}}\else{$f_{\rm gas}$}\fi}
\newcommand{\cubecm}{\ifmmode{{\rm cm^{-3}}}\else{cm$^{-3}$}\fi}
\newcommand{\ztwo}{\ifmmode{{\rm [Z_2/H]}}\else{[Z$_2$/H]}\fi}
\newcommand{\zthree}{\ifmmode{{\rm [Z_3/H]}}\else{[Z$_3$/H]}\fi}
\newcommand{\lsim}{\lower0.3em\hbox{$\,\buildrel <\over\sim\,$}}
\newcommand{\gsim}{\lower0.3em\hbox{$\,\buildrel >\over\sim\,$}}

\newcommand{\sfr}{\ifmmode{\textrm{M}_\odot \,\textrm{yr}^{-1} \,\textrm{Mpc}^{-3}}\else{M$_\odot$ yr$^{-1}$ Mpc$^{-3}$}\fi}
\newcommand{\hsfr}{\ifmmode{\textrm{M}_\odot\, \textrm{yr}^{-1}}\else{M$_\odot$ yr$^{-1}$}\fi}

\newcommand{\eavg}{\ifmmode{\langle E_\gamma \rangle}\else{$\langle E_\gamma \rangle$}\fi}

\newcommand{\enzo}{{\sc enzo}}
\newcommand{\grackle}{{\sc grackle}}
\newcommand{\yt}{{\sc yt}}
\newcommand{\ytree}{{\sc ytree}}

\newcommand{\Ms}{\ifmmode{M_\odot}\else{$M_\odot$}\fi}
\newcommand{\vrms}{\ifmmode{v_{\rm rms}}\else{$v_{\rm rms}$}\fi}

\newcommand{\tvir}{\ifmmode{T_{\rm{vir}}}\else{$T_{\rm{vir}}$}\fi}
\newcommand{\mvir}{\ifmmode{M_{\rm{vir}}}\else{$M_{\rm{vir}}$}\fi}
\newcommand{\rvir}{\ifmmode{r_{\rm{vir}}}\else{$r_{\rm{vir}}$}\fi}

\newcommand{\jj}{\ifmmode{J_{21}}\else{$J_{21}$}\fi}
\newcommand{\flw}{\ifmmode{F_{LW}}\else{$F_{LW}$}\fi}
\newcommand{\kph}{\ifmmode{k_{\rm ph}}\else{$k_{\rm ph}$}\fi}

\newcommand{\zsun}{\ifmmode{\rm\,Z_\odot}\else{$\rm\,Z_\odot$}\fi}

\newcommand{\nhi}{\ifmmode{N_{\rm HI}}\else{$N_{\rm HI}$}\fi}

\newcommand{\rockstar}{{\sc Rockstar}}

\def\eps@scaling{1.0}%
\newcommand\epsscale[1]{\gdef\eps@scaling{#1}}%
\newcommand\plotone[1]{%
 \centering 
 \leavevmode 
 \includegraphics[width={\eps@scaling\columnwidth}]{#1}%
}%
\newcommand\plottwo[2]{%
 \centering 
 \includegraphics[width={\eps@scaling\columnwidth}]{#1}%
 \hfil 
 \includegraphics[width={\eps@scaling\columnwidth}]{#2}%
}%

\makeatletter
\newcommand*{\rom}[1]{\expandafter\@slowromancap\romannumeral #1@}
\makeatother

\title[Neutron Star Mergers from Population III Stars]{Neutron Star Mergers and their Impact on Second Generation Star Formation in the Early Universe}

\author[Skinner \& Wise]{
Danielle Skinner$^{\orcidlink{0000-0002-5346-1308}}$$^{1}$\thanks{E-mail: drenniks@gatech.edu},
John H. Wise$^{\orcidlink{0000-0003-1173-8847}}$$^{2}$
\\
$^{1}$Department of Physics, Oregon State University, 301 Weniger Hall, Corvallis, OR 97331, USA\\
$^{2}$Center for Relativistic Astrophysics, School of Physics, Georgia Institute of Technology, Atlanta, GA 30332, USA\\
}

\date{Accepted XXX. Received YYY; in original form ZZZ}

\pubyear{2024}

\begin{document}
\label{firstpage}
\pagerange{\pageref{firstpage}--\pageref{lastpage}}
\maketitle

\begin{abstract}
	The exact evolution of elements in the universe, from primordial to heavier elements produced via the r-process, is still under scrutiny. The supernova deaths of the very first stars led to the enrichment of their local environments, and can leave behind neutron stars (NS) as remnants. These remnants can end up in binary systems with other NSs, and eventually merge, allowing for the r-process to occur. We study the scenario where a single NS merger (NSM) enriches a halo early in its evolution to understand the impact on the second generation of stars and their metal abundances. We perform a suite of high resolution cosmological zoom-in simulations using \enzo{} where we have implemented a new NSM model varying the explosion energy and the delay time. In general, a NSM leads to significant r-process enhancement in the second generation of stars in a galaxy with a stellar mass of $\sim 10^5 \Ms$ at redshift 10.  A high explosion energy leads to a Pop II mass fraction of 72\% being highly enhanced with r-process elements, while a lower explosion energy leads to 80\% being enhanced, but only 14\% being highly enhanced. When the NSM has a short delay time of 10 Myr, only 5\% of the mass fraction of Pop II stars is highly enhanced, while 64\% is highly enhanced for the longest delay time of 100 Myr. This work represents a stepping stone towards understanding how NSMs impact their environments and metal abundances of descendant generations of stars.
\end{abstract}

\begin{keywords}
stars: Population III -- stars: Population II -- transients: neutron star mergers -- methods: numerical
\end{keywords}



\section{Introduction}

Nucleosynthesis in the universe is a complex accounting task to undertake. Determining the source of each element of the periodic table takes great attention to the details of the particle physics and chemistry that underlies these processes. The heaviest elements of the periodic table are produced by the rapid neutron capture process, called the r-process. During this process, in regions of very high neutron densities, when the rapid bombardment of neutrons onto nuclei is faster than the rate of beta decay, nuclei can build up to produce heavy, stable elements, like Eu, Os, Ir, Pt, Th, and U. But the exact origin of this process hasn't been completely pinned down yet \citep{Kajino19, Cowan21}. 

The r-process likely takes place in events that are either related to massive stars or through compact binary mergers. SNe of massive stars can produce r-process elements \citep[see][for a discussion on different sources]{Cowan21}. Core collapse SNe (CCSN) explosions ignited by the magnetorotational explosion model produce r-process elements in both the weak and the strong magnetic field models, with preference given to the strong magnetic field models to explain some r-process abundances of metal-poor stars \citep{Nishimura15}. Collapsars, rapidly rotating massive stars that undergo CCSN, can also produce r-process material \citep{Barnes22}. \citet{Brauer21} create a model of a population of collapsars to study these objects as the source of r-process abundances in metal-poor stars. They find that collapsars can explain the Eu abundances within metal-poor stars, although they cannot rule out NSMs as an alternative or dominant source.

In this work, we are concerned with compact binary mergers, and specifically with two NSs that merge together. Observationally, this has been confirmed. The GW170817 event was detected by LIGO, and along with it came a kilonova, a transient burst of energy powered by radioactive decay, making this event a multimessenger astronomical affair. LIGO detected this NSM with components in the mass range of $1.17 - 1.60 \Ms$ \citep{Abbott17}, and a $\gamma$-ray burst 1.7 seconds after the merger was detected by the Fermi Gamma-ray Burst Monitor \citep{Goldstein17}. This allowed for precise sky localization and a massive follow up effort from telescopes across the electromagnetic spectrum \citep[see][for an extensive review of this follow-up]{Abbott17a}. For the first time, we not only observed a NSM via GWs, but we also confirmed the association between NSMs and $\gamma$-ray bursts. Out of this massive follow-up campaign came analysis of the light curve of the kilonova in the UV, optical, and IR \citep{Drout17, Pian17}. \citet{Drout17} found that this kilonova was in fact powered by radioactive decay of r-process elements. They estimate that $\sim 0.05 \Ms$ of enriched, r-process material must have been ejected from this event. This was the first tangible evidence that NSMs can produce a significant amount of r-process material.

Studies of the abundance of r-process elements that NSMs can produce has shown that these events can explain the r-process abundances on a global and local scale. \citet{Goriely11} used relativistic smoothed particle hydrodynamics simulations to study the ejecta of a NSM scenario. They model the merging of an equal mass NS system, $1.35:1.35 \Ms$, and an asymmetric system, $1.2:1.5 \Ms$ and use two different EoSs, one representing a ``soft'' EoS and the other a ``stiff'' EoS. A total mass yield of $\sim 10^{-3} - 10^{-2} \Ms$ is ejected in these scenarios, with the asymmetric model ejecting more material than the symmetric model. They find that nearly all galactic Eu could be explained by the yields they determined, making NSMs a promising dominant source of r-process elements. 

A piece of indirect evidence comes from studying r-process abundances in metal-poor stars. \citet{Tarumi21} studied the Ba abundance, an r-process tracer element, within metal-poor stars to constrain the delay time between the formation of the binary system, and the production of r-process elements, via a NSM. In simple terms, they are investigating to see if the timing works out for the r-process enriched material to end up in metal-poor stars we see today, given their abundances. They find that as the [Fe/H] abundances increase, so do the [Ba/Mg] abundances, implicating a time delay of about 100 Myr to 1 Gyr in the r-process enrichment. \citet{Tarumi21} confidently declare that the origin of r-process elements must be sources with a time-delay baked in to the model. More locally, \citet{Bartos19} studied daughter isotopes of r-process elements found in meteorites in our solar system. Using numerical simulations of abundances in the early Solar System, they find that a NSM may have occurred $\sim 300$ pc away, 80 Myr before the formation of the Solar System, and deposited its r-process material into the pre-Solar System environment. The case for NSMs grows stronger.

The motivation for this work comes from the r-process enriched stars in Reticulum II (Ret II), an ultrafaint dwarf (UFD) galaxy which serves as an ideal laboratory for studying early universe star formation. These systems typically form most of their stars early in the universe, before reionization was able to suppress star formation \citep{Weisz14}. This means that these stars form in the first couple hundred million years after the Big Bang and haven't been affected by recent mergers or enriched by recent star formation. Ret II was detected by the Dark Energy Survey \citep{Bechtol15}, whose stars have since been extensively studied \citep{Simon15, Roederer16, Ji16, Ji22}. Recently, \citet{Ji22} have found that $\sim 72\%$ of the stars in Ret II are r-process enriched, whose source could have been a prompt NSM. Since this galaxy is so isolated, it serves as an ideal case study for simulating this scenario: namely, what happens if a single NSM, produced from the remnants of Pop III stars, occurs within a galaxy in the early universe? This type of scenario is particularly unique due to the chemical impact a NSM could have on the star formation history of a galaxy. If a NSM did occur early in the history of Ret II, the r-process enriched stars are a special and direct connection to previous generations of stars.

This paper represents the proof of concept for a new NSM model that we have incorporated into \enzo{}, an adaptive mesh refinement (AMR) cosmological simulation code. We run high-resolution cosmological simulations of a minihalo from a redshift of $z = 250$ to $z = 10$, varying the explosion energy and delay time of a NSM. We follow the material ejected into the local environment from this event into the next generation of stars to understand how this material is mixed within the environment and affects the next population of stars. The paper is structured as follows: the methods of the simulation is discussion in \S \ref{sec:methods}, with the added NSM model and the simulation suite described in \S \ref{sec:NSM_model}. Results are presented in \S \ref{sec:results}, followed by a discussion in \S \ref{sec:discussion}. We conclude with \S \ref{sec:conclusion}.

\section{Methods} \label{sec:methods}

\subsection{Simulation set-up} \label{sec:NSM_simulation}

We use the AMR, hydrodynamical simulation code \enzo{} \citep{Enzo} to run high-resolution proof-of-concept cosmological simulations of a NSM within a zoom-in simulation of a single progenitor halo, and we analyze the simulations with the analysis toolkit \yt{} \citep{yt_full_paper}. We run a total of six simulations, five with a NSM model and one without. The run without the NSM model was run first, and the following five NSM simulations were run based on the non-NSM (original) simulation. The original simulation has a domain width of 1 comoving Mpc and a root grid of $64^{3}$ cells. We use the 12-species (H$_{2}$, H$_{2}^{+}$, H$^{-}$, H, H$^{+}$, He, He$^{+}$, He$^{++}$, e$^{-}$, D, D$^{+}$, and HD) non-equilibrium chemistry model \citep{Abel97, Anninos97}. We consider metal cooling and use the chemistry and cooling library \grackle{} \citep{grackle}. We run this simulation to $z = 10$ and then choose the most massive halo in the box to be the target of a zoom-in simulation. We choose a Lagrangian volume around this halo, and re-initialize the simulation with two more nested grids. The highest refinement level of $l = 14$ has a minimum cell size of $8.6 \times 10^{-2}$ pc and a dark matter mass resolution of $1985 \Ms$. We refine a cell when the baryon or dark matter overdensity exceeds $3 \bar{\rho}_{\mathrm{b, DM}} \times 2^{l (3 + \epsilon)}$ where $l$ is the refinement level and $\epsilon = -0.3$ makes the refinement super-Lagrangian. We also refine around must-refine particles that are initialized in the halo's Lagrangian region, which allows for added refinement in the 8 nearest cells around moving regions up to a level of two. The Jeans length is always resolved by at least four cells. We use the cosmological parameters from the Planck Collaboration (2020) \citep{Planck18_Cosmo}: $\Omega_{\mathrm{M}} = 0.3111$, $\Omega_{\Lambda} = 0.6889$, $\Omega_{\mathrm{DM}} = 0.262125$, $\Omega_{\mathrm{b}} = 0.04897$ and $h = 0.6766$, with the parameters defined as their usual definitions.

Once the original simulation finished, we restarted the simulation with different NSM parameters. Specifically, we vary the explosion energy of the NSM itself and the delay time, the time between NS binary system formation and NSM. In the next subsections, we discuss the NSM model that has been added to \enzo{}, and the NSM simulation suite.

\subsection{Neutron star merger model and simulation suite} \label{sec:NSM_model}
To explore how a NSM descendent from Pop III stars affects the r-process enrichment of Pop II stars, we implement a new NSM model within \enzo{} to control the delay time and the explosion energy of the event, modelled as a single star particle. Upon completion of the original simulation, a single Pop III star is chosen as the candidate NSM event. The chosen Pop III star must be between $20 \leq M / M_{\odot} \leq 30$ so that when this system is split into a symmetric binary system, the initial masses of the component Pop III stars is between $10 \leq M / M_{\odot} \leq 15$, corresponding to the SN models from \citet{Heger10} which will produce a NS as a remnant. We use their model C as a point of reference since the explosion energy, $0.9 \times 10^{51}$ erg, is the closest to the CCSN explosion energy in \enzo{}. The simulation is restarted at the output where the Pop III star forms, indicating which Pop III star and what NSM parameters should be included. We chose the first allowed Pop III star that forms in the simulation to become the source of the NSM event in order to allow for the maximum amount of time between the NSM event and for the second generation of stars to form. The Pop III star that we chose has an initial mass of $28.7 M_{\odot}$ that was randomly selected from an IMF described in \citet{Skinner20}. The chosen Pop III star is then converted into a symmetric Pop III binary system with component masses of $\sim 14.35 M_{\odot}$, but is still modelled as a single star particle with an ionizing hydrogen photon rate of $2.9 \times 10^{48} \mathrm{s}^{-1}$, which is equal to the sum of the two individual stellar photon rates from \citet{Schaerer02}.

The Pop III binary system then undergoes its first explosion via a CCSN with an explosion energy of double that of a normal CCSN since we are modelling two stars in one star particle, with a total explosion energy of $2 \times 10^{51}$ erg. The remnant particle is now labelled as a NS binary system. To determine the masses of the remnant NSs, we fit a parabola (Equation \ref{eq:nsm_mass}) to the model C data from Figure 6 in \citet{Heger10}. The coefficients are fit to within 1\% of the standard error. The remnant now has a total mass of $\sim 4.8 M_{\odot}$, with the component masses each having a mass of $\sim 2.4 M_{\odot}$, making this system a higher mass NS binary system.

\begin{equation} \label{eq:nsm_mass}
	M_{\mathrm{NS}} = 0.01319 M_{\mathrm{PopIII}}^{2} - 0.0421 M_{\mathrm{PopIII}} + 0.301
\end{equation}

A second explosion goes off with $E_{\textrm{NSM}}$ after a time $t_{\textrm{merge}}$, with the parameters chosen in a particular run given the choices in Table \ref{tab:NSM_params}. The final remnant particle is then converted to an inert particle with a mass equal to the sum of the remnant masses. We do not consider any EM radiation from the NSs nor any kilonova effects of a NSM for simplicity; we only follow the mass ejected from this event. We track the r-process material in a passive scalar field which does not affect metal cooling rates. For this proof of concept, the total ejecta mass is set to $0.01 \Ms$, but in practice, due to the Cartesian discretization of a sphere in of Enzo, the actual mass ejected is $0.0169 M_{\odot}$. The delay times chosen for these runs are on the shorter end of timescales for the merging of two NSs to take place, but for this proof of concept, we explore this particular situation where a NSM occurs early in the star formation history of the galaxy. For a galaxy like Ret II, a NSM would have had to occur within 500 Myr to allow for ample time for thorough mixing to take place within the halo \citep{Simon23}. Delay times of 10, 100, and 300 Myr are short, but are not completely out of the realm of possibilities, and allow us to see how a NSM affects subsequent star formation at this early point in time. Short delay times for these systems may be possible, yet rare, if the system has a low natal kick and undergoes a common envelope phase. Through this fast-formation channel, a shorter delay time has been studied with minimum delay times ranging from 0.1 Myr to 10 Myr \citep{Belczynski01, Dominik12, Belczynski18,Jeon21}.

Table \ref{tab:NSM_params} shows the NSM parameters for each run of this simulation suite. Each simulation was restarted when the chosen Pop III star first forms, at $z = 20.1$ and is run until $z = 10$. Before $z = 20.1$, all simulations share the same datasets as the original run. We run \rockstar{} \citep{rockstar} on each run to identify the main halo and use \ytree{} \citep{ytree} to determine the most massive progenitor line.

\begin{table}
	\centering
	\begin{tabular}{lcc} 
		\hline
		Simulation & $t_{\textrm{merge}}$ [Myr] & E$_{\textrm{NSM}}$ [erg] \\
		\hline
		Fiducial (C) & 100 & $10^{50.5}$ \\
		A & \textbf{10} & $10^{50.5}$ \\
		B & \textbf{30} & $10^{50.5}$ \\
		D & 100 & $\mathbf{10^{50}}$ \\
		E & 100 & $\mathbf{10^{51}}$ \\
		No NSM Event & -- & -- \\
		\hline
	\end{tabular}
	\caption{Parameter choices for each simulation. Non-Fiducial parameters are bolded.}
	\label{tab:NSM_params}
\end{table}

\subsection{Pop III and Pop II star formation and feedback} \label{sec:star_formation}
We will briefly discuss the formation of Pop III and Pop II stars in the simulation. For more detailed descriptions, we refer the reader to \citet{Skinner20}. A Pop III star will form when the following criteria are met within a cell: 
\begin{enumerate}
	\item metallicity of $\mathrm{Z} \leq 1.295 \times 10^{-6} \mathrm{Z}_{\odot}$
	\item gas number density of $\mathrm{n} > 10^{6} \mathrm{cm}^{-3}$
	\item converging gas flow: $\nabla \cdot \mathrm{v}_{\mathrm{gas}} < 0$
	\item molecular hydrogen number fraction of $\mathrm{f}_{\mathrm{H}_{2}} > 10^{-3}$
\end{enumerate}
Pop III stars have mass-dependent lifetimes and produce hydrogen ionizing and Lyman-Werner luminosities from \citet{Schaerer02}. Their SN and end states depend on their mass-dependent stellar endpoints. A Pop III particle that does not undergo a NSM represents a single Pop III star. 

A Pop II star forms using the same criteria, except without the molecular hydrogen fraction requirement, and with a metallicity greater than the above value. A Pop II particle represents a cluster of Pop II stars with a minimum cluster mass of 1000 \Ms. There are instances when a Pop II star will form below the minimum mass, in which case the star accretes mass without producing feedback until it reaches that minimum mass. Both Pop III and Pop II stars are assigned a metallicity based on the metallicity fraction of the cell that the particle forms in. Because Pop II stars form as star clusters, the exact number of Pop II star particles that form is not a good indicator of how productive an environment is at forming Pop II stars. Rather, the total Pop II stellar mass is the indicator that will be presented throughout this work. We do not take into account the effects of Pop III or Pop II binary systems due to the limiting resolution of our simulations.

\section{Results} \label{sec:results}

We discuss the evolution of the main halo in the original run in \S \ref{sec:main_halo}. In \S \ref{sec:energy_vary} and \S \ref{sec:time_vary}, we describe the explosion energy variation runs and the delay time variation runs. When we discuss the r-process metallicities, we are only considering r-process material from the NSM. In this sense, we are presenting the lower-bound of r-process material in our system as we are not accounting for any r-process material produced via core-collapse SNe from Pop III and Pop II stars.

\subsection{Evolution of the main halo in the original run} \label{sec:main_halo}
\begin{figure} 
	\includegraphics[width=\columnwidth]{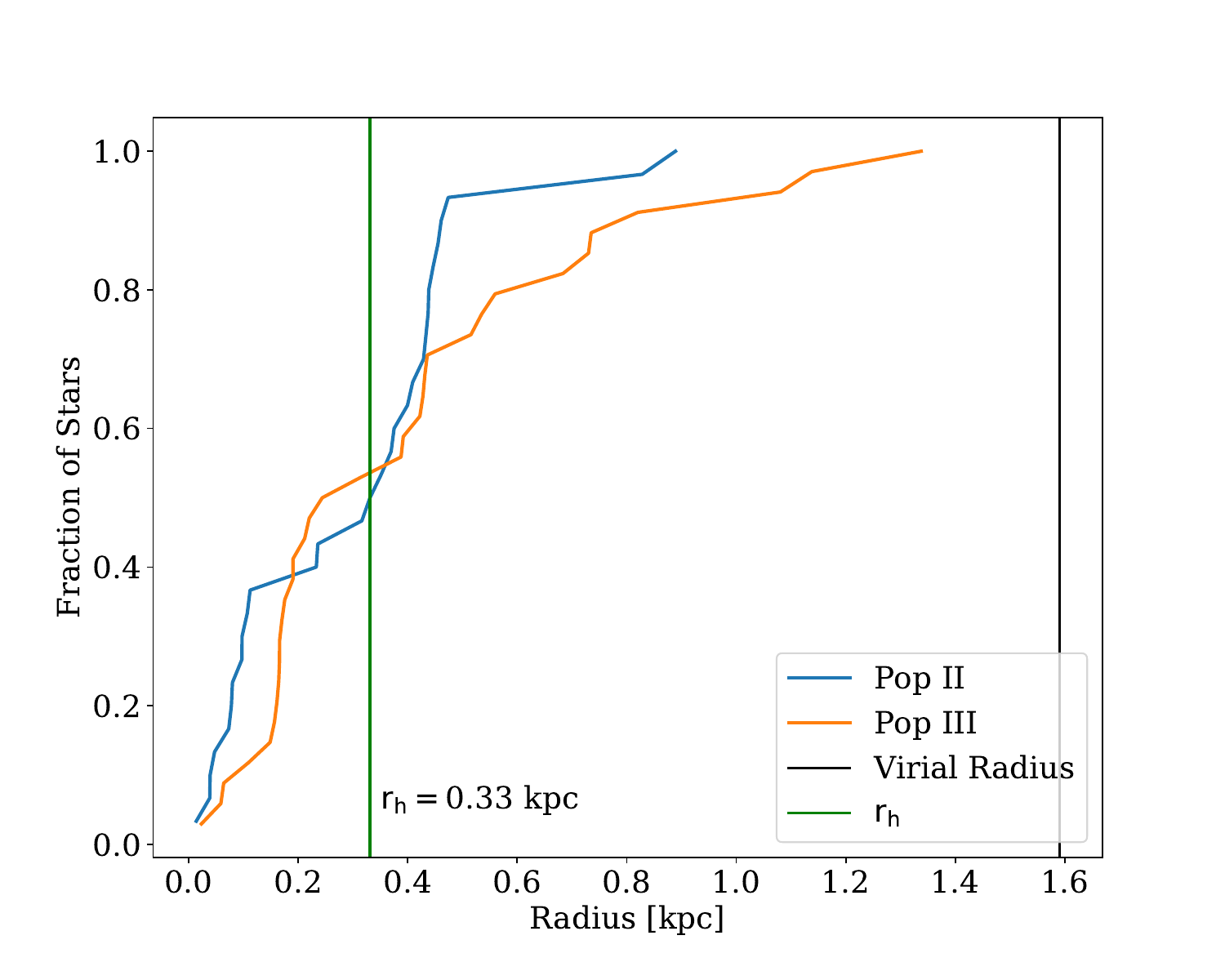}
	\caption[Radial distribution of Pop II and III stars in the main halo.]{The radial distribution of Pop III and Pop II stars in the halo in the original run at $z = 10$. The orange and blue lines show the Pop III and Pop II stars, respectively. The black line indicates the virial radius of the halo. The half-mass radius for this halo is 331.2 pc}
	\label{fig:og_stellarmass}
\end{figure}

\begin{figure*} 
	\includegraphics[width=\textwidth]{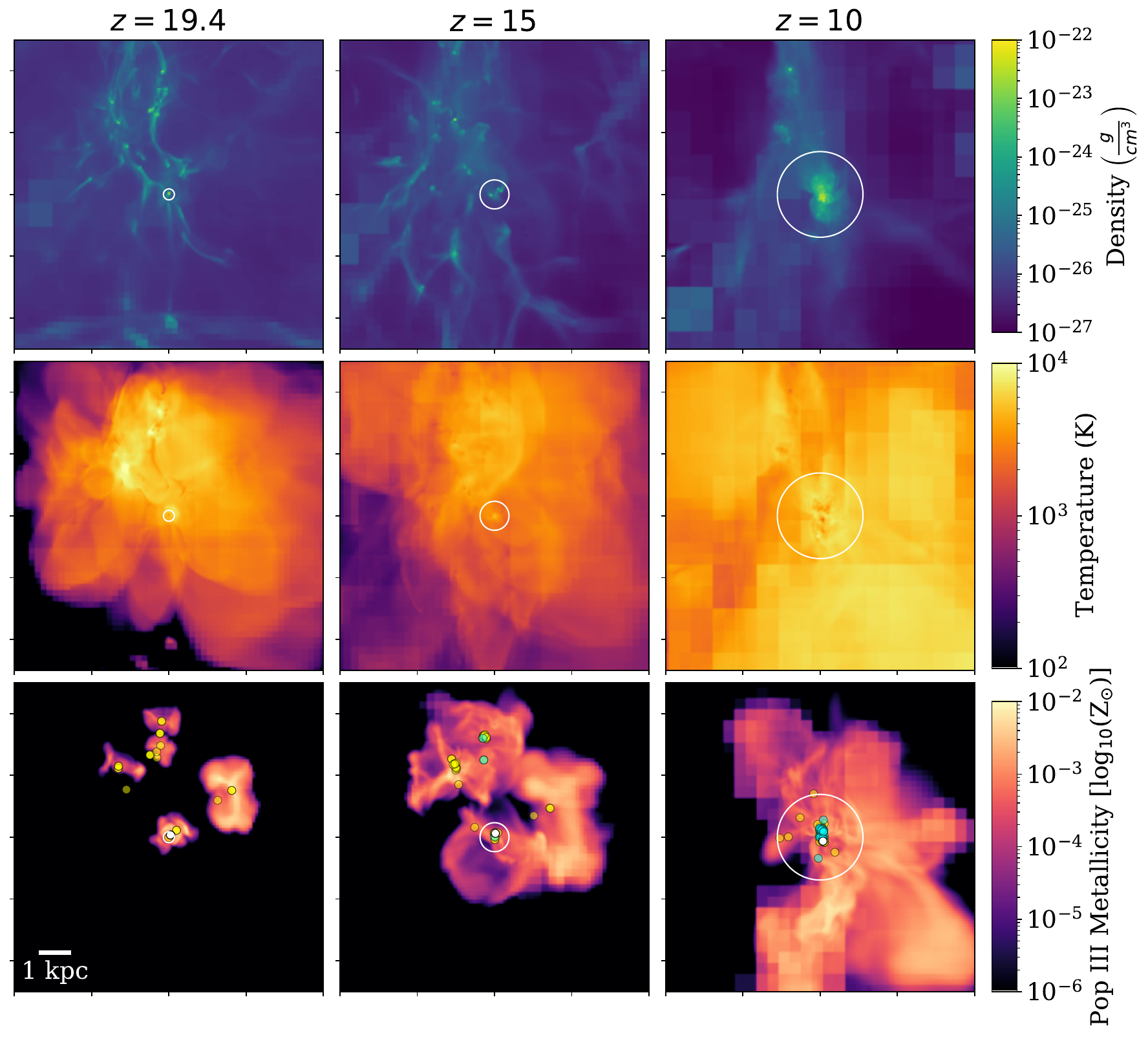}
	\caption[Projections of the density, temperature and Pop III metallicity of the main halo through time.]{Evolution of the main halo. The left column shows the $z = 19.4$ timestep, when the chosen Pop III star explodes as a hypernova, the middle $z = 15$, and the right $z = 10$. From top to bottom, the rows show the density-weighted projections of gas density, temperature, and metallicity from Pop III stars. The star particles are only plotted in the bottom row. The white circle indicates the halo virial radius, the yellow dots show the Pop III stars, and the aqua dots show the Pop II stars. The white dot indicates the particle chosen to be the NSM in the other runs.}
	\label{fig:og_evolution}
\end{figure*}

In the original run with no NSM, the most massive progenitor of the main halo is first identified at $z = 26.9$ with a mass of $10^{4.6} \Ms$, and ends with a mass of $10^{8.19} \Ms$ at $z = 10$. During that time, the halo forms two Pop III stars with masses of 26.5 \Ms{} and 195.5 \Ms{} at 150.7 Myr ($z = 22.6$) and 286.1 Myr ($z = 14.4$), respectively. The halo experiences 25 Pop III SN from stars that accrete into the halo and from the two that form within the main progenitor of the halo. 26 Pop II star particles form within the halo with a total stellar mass of $10^{4.5} \Ms$, and with a median and mean mass weighted creation time of 441 Myr ($z = 10.5$) and 402 Myr ($z = 11.3$), respectively. There are also five Pop II star particles that fall inside the halo. Figure \ref{fig:og_stellarmass} shows the radial distribution of Pop III and Pop II stars within the halo at $z = 10$. The half-mass radius is 331.2 pc, which is on the higher end of results from other simulations of UFD galaxies \citep{Ricotti16}. The Pop II stars all tend to cluster within 1 kpc of the center of the halo, while the Pop III stars are more dispersed throughout the halo. 

This halo merges with a neighboring smaller halo that has formed its own distribution of Pop III and II stars, which is where a majority of the stars that fell in came from. This results in a mix of stars that reside within the final halo, those that formed within the main progenitor, and those that formed outside. This can be seen in Figure \ref{fig:og_evolution}, where Pop III and Pop II stars form within 5 kpc of the main halo, and end up combining within the final halo at $z = 10$. This feature will be of particular interest in the NSM runs. 

By $z = 10$, the halo has a Pop II stellar mass of $\sim 10^{4.4} \Ms$ and a living Pop III stellar mass of $\sim 978 \Ms$. At this time, there are still 14 Pop III stars living within the halo, concurrently with Pop II stars. We stress that the number of Pop II star particles is not of particular importance, since they represent clusters of Pop II stars. The overall mass distribution of Pop II stars is relevant, and thus when discussing the Pop II distribution, we will focus on the Pop II stellar masses. 

\subsection{Explosion energy variation runs} \label{sec:energy_vary}

In each of these runs, where the NSM explosion energy is varied in the range $10^{50 - 51}$ erg, the NSM takes place at the same time at $286$ Myr ($z = 14.4$). The general dynamics of the halo remain unchanged, and the main halos in the NSM runs looks visually similar to the halo in the original run. Run D ($10^{50}$ erg), Fiducial ($10^{50.5}$ erg), and E ($10^{51}$ erg) have Pop II stellar masses of $10^{5.27}$, $10^{4.87}$, and $10^{4.80}$ \Ms{}, at $z = 10$ respectively, i.e. the lowest energy run produces the highest Pop II stellar mass. The median and mean mass weighted creation times for the Pop II stars that form within the halo for runs D, Fiducial, and E are 462 Myr ($z = 10.2$) and 455 Myr ($z = 10.3$), 456 Myr ($z = 10.3$) and 443 Myr ($z = 10.5$), and 443 Myr ($z = 10.5$) and 436 Myr ($z = 10.6$) respectively. The mean creation time for Pop II stars in these runs is higher than the mean creation time of 402 Myr in the original run. In the main progenitor, runs D and Fiducial form two Pop III stars at 151 Myr ($z = 22.6$) and 353 Myr ($z = 12.4$) and at 151 Myr ($z = 22.6$) and 266 Myr ($z = 15.2$), with masses of 26.5 \Ms{} and 27.7 \Ms{}, and 26.5 \Ms{} and 30.4 \Ms, respectively. Run E produces only one Pop III star at 151 Myr ($z = 22.6$) with a mass of 26.5 \Ms{}. The halos in runs D, Fiducial and E experience 39, 38, and 33 Pop III SNe and contain 22, 23, and 19 living Pop III stars at $z = 10$.

\begin{figure} 
	\includegraphics[width=\columnwidth]{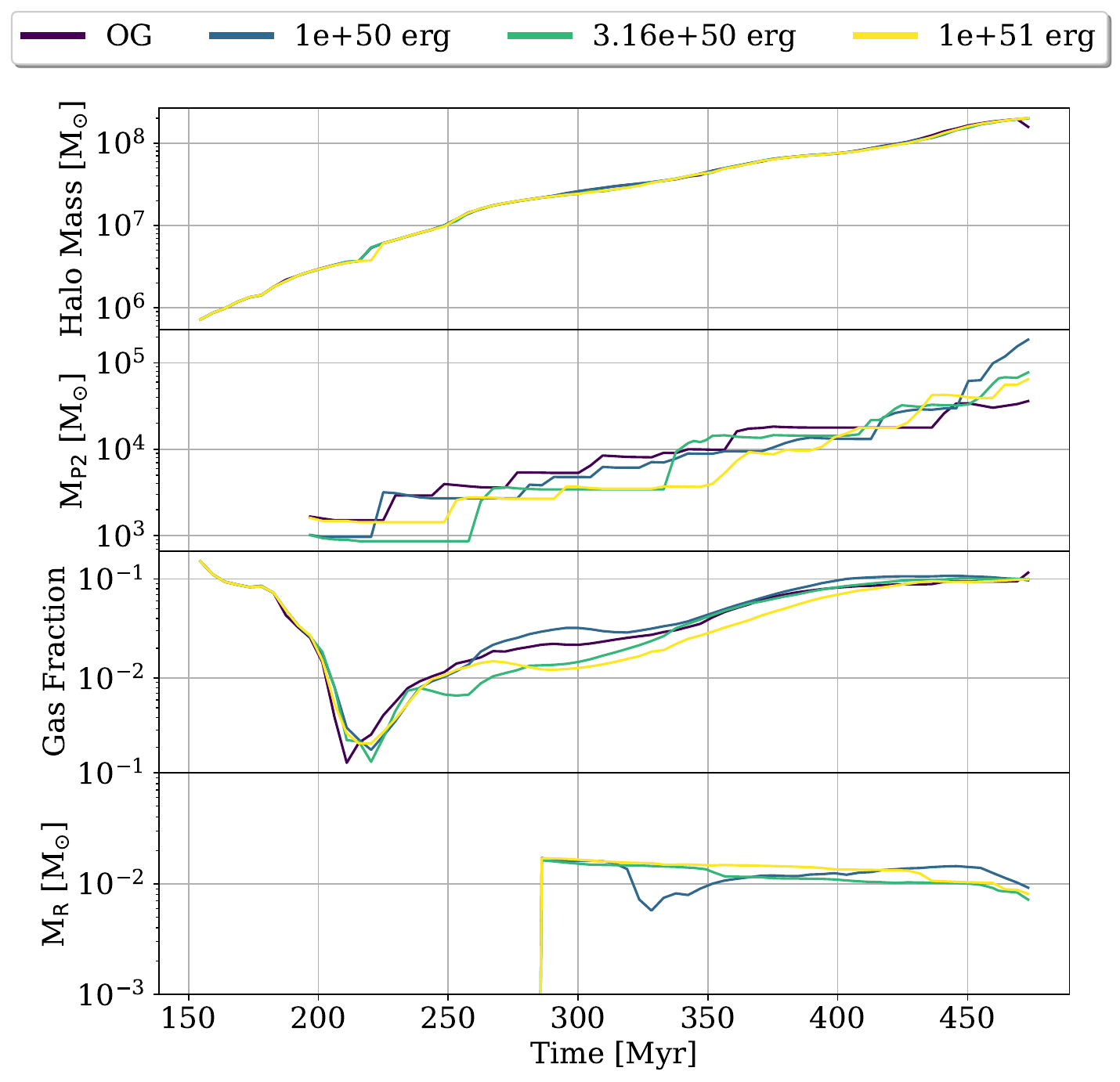}
	\caption[The halo mass, Pop II stellar mass, gas fraction, and r-process mass as a function of time for the explosion energy variation runs.]{The evolution of the main halo in each explosion energy variation run and the original (OG) run. The top panel shows the halo mass, second from the top shows the Pop II stellar mass, third from the top shows the average gas fraction, and the bottom shows the r-process mass. The NSM occurs at 286 Myr.}
	\label{fig:eng_evol}
\end{figure}

We show the halo mass, Pop II stellar mass, gas fraction, and r-process mass of the most massive halo progenitor in Figure \ref{fig:eng_evol}. As expected, halo masses in each run are almost identical to each other, and with runs D, Fiducial, and E growing to $10^{8.3} \Ms$ and the original run to $10^{8.2} \Ms$. The Pop II stellar mass follows a relatively similar trend in each run, but the NSM runs end with larger Pop II stellar masses as compared to the original run. The lowest energy run forms the most, followed by the middle energy, and finally the highest energy run. This is consistent with what is expected of an explosion within a halo. The higher the explosion energy, the more vulnerable the halo will be to disruption from the event. The gas fraction dips at around 200 Myr in each run due to the formation of the first stars within the halo, which is quite small at this point in time. This star formation event blows out most of the gas through radiative and SN feedback. As halo mergers takes place, the gas fraction recovers and steadily rises to $\sim 0.1$. The bottom panel of Figure \ref{fig:eng_evol} shows the r-process mass ejected from the NSM within the halo as a function of time. When the NSM occurs, the entirety of the explosion is contained within each halo. The lowest energy run experiences an abrupt dip in the r-process mass due to some of the mass being blown out by SN that occur outside and within the halo. This mass is eventually gravitationally pulled back into the halo as merging occurs. The r-process mass in each run slowly decreases as star formation takes place and some of the r-process material is deposited into star particles. 
\begin{figure*} 
	\includegraphics[width=\textwidth]{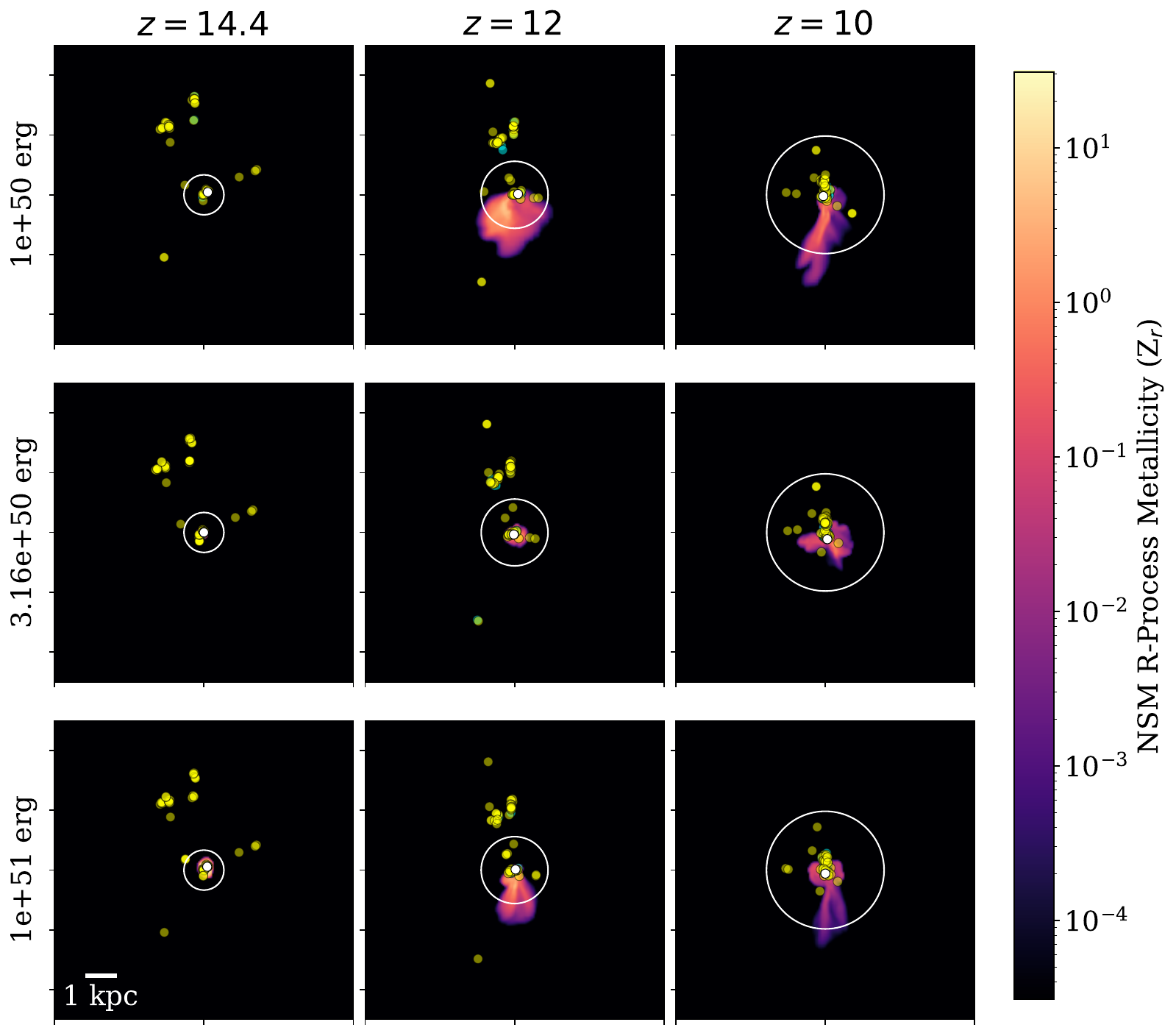}
	\caption[R-process metallicity projections of the explosion energy variation runs through time.]{Evolution of the density-weighted projections of the r-process metallicity for each explosion energy variation run. The first column shows the system at $z = 14.4$, when the NSM event takes place, the middle column shows $z = 12$, and the right column shows the final output at $z = 10$. The top row shows run D, the middle shows the Fiducial run, and the bottom shows run E. The white circle indicates the virial radius of the halo. The yellow dots show Pop III stars, and the aqua dots show Pop II stars. The white dot indicates the chosen Pop III star that produced the NSM event. Note that the star particles are covering the NSM explosion in the lowest and Fiducial runs at the first timestep.}
	\label{fig:eng_met4_evol}
\end{figure*}
The evolution of the r-process metallicity field can be seen in Figure \ref{fig:eng_met4_evol}. The first column shows the timestep where the NSM takes place. At this time, the highest energy run expands more quickly into the halo, as expected. Later at $z = 12$, the Fiducial run (middle energy) is still fully contained within the halo, while the lowest and highest energy runs are starting to escape. This is caused by a SN that occurred just outside the halo and within the halo right around $z = 12$ in the lowest and higher energy runs. This blows some of the r-process enhanced gas out of the halo in these runs. By the end of the simulation at $z = 10$, any r-process enhanced gas that was lost due to stellar feedback has mostly returned to the halo. This is also evident in the bottom panel of Figure \ref{fig:eng_evol}, where the r-process mass in the lowest energy run dips at around $\sim 325$ Myr, but recovers quickly as that gas gravitates back inwards. This effect is most prominent in the lower energy run. All runs have a slight decrease in the r-process mass with time and as star formation occurs and consumes it.

To better represent the r-process metallicity, we define a new unit $Z_{\mathrm{r}} = 0.01384$ which is scaled to the r-process abundances of the solar system from \citet{Goriely99}, normalized by the solar system Si abundance. This means that a metallicity of 1 $Z_{\mathrm{r}}$ is equal to the r-process abundances within the solar system. To convert from $Z_{\mathrm{r}}$ to an abundance ratio of [X/H] for some element X, one can use $[X/H] = [X/H]_{\mathrm{r}} + A(X)_{\odot, r} - A(X)_{\odot}$ to normalize to an r-process solar composition \cite[see Step \#3 in Table 2 of][for an excellent description of metallicity conversions]{Hinkel22}. Here, $[X/H]_{\mathrm{r}} = v_{x} + \mathrm{log}(Z_{\mathrm{r}})$, where $v_{x}$ is open to the reader, depending on their choice of NSM ejecta yields. This makes $[X/H] = v_{x} + \mathrm{log}(Z_{\mathrm{r}}) + A(X)_{\odot, r} - A(X)_{\odot}$. To compare with results from \citet{Frebel23}, we put this in terms of abundances from \citet{Lodders09} (L09) and \citet{Roederer22} (R22) as collected by \citet{Frebel23} (see their Table 2). In terms of these values, $[X/H] = v_{x} + \mathrm{log}(Z_{\mathrm{r}}) + \mathrm{log} \epsilon_{\mathrm{R22}} (X) - \mathrm{log} \epsilon_{\mathrm{L09}} (X)$. For Ba, the difference between the abundances from \citet{Frebel23} is --0.92 and for Eu, the difference between the abundances is --0.12. 

\begin{figure} 
	\includegraphics[width=\columnwidth]{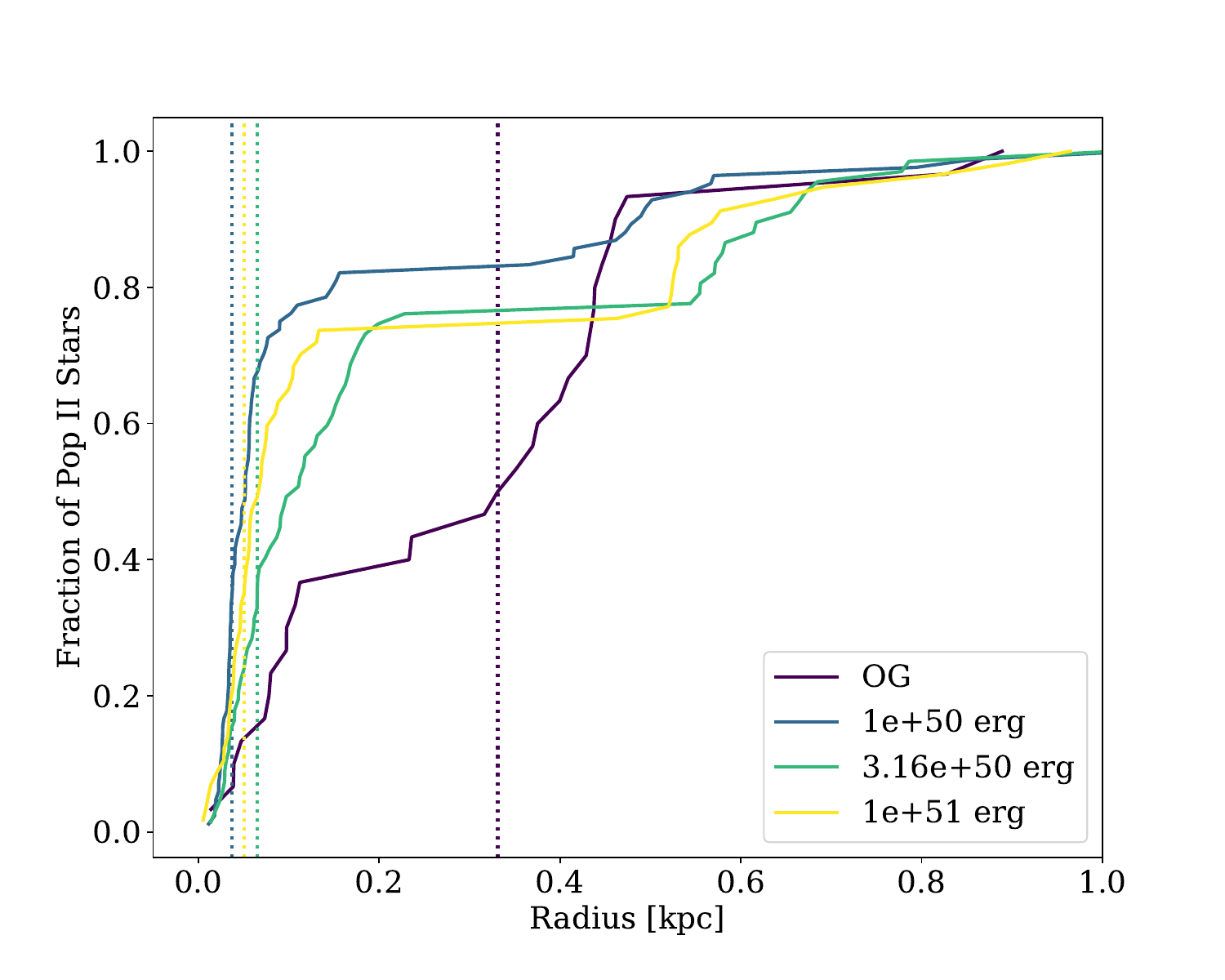}
	\caption[Radial distribution of Pop II stars for each explosion energy variation run.]{The fraction of Pop II stars that are within some radius at the $z = 10$ for runs D, Fiducial and E, and the original run. The dashed lines indicate the half-mass radius, which are 331.2 pc, 37.1 pc, 65.3 pc, and 50.6 pc for the original run, run D, run Fiducial, and run E, respectively. The virial radius at this time is 1.6 kpc, 1.73 kpc, 1.72 kpc, and 1.73 kpc for the original run, run D, run Fiducial, and run E, respectively.}
	\label{fig:eng_stellar_dist}
\end{figure}

The energy variation runs lead to a significantly higher Pop II stellar masses as compared to the original run. Table \ref{tab:eng_stellar} shows the number of Pop III stars and the total Pop II stellar mass in the first two columns. A group of Pop II stars form and cluster within the dense central region of the halo where the NSM takes place, in each run. Figure \ref{fig:eng_stellar_dist} shows the fraction of Pop II stars within some radius for each run. The dotted lines show the half-mass radius, which is much smaller for the NSM runs, with half-mass radii of 37.1 pc, 65.3 pc, and 50.6 pc for runs D, Fiducial, and E. Compared to the original run whose half-mass radius is 331.2 pc, there is a much larger fraction of Pop II stars that have clustered in the central 0.2 kpc, leading to the lower half-mass radii. These smaller half-mass radii are more aligned with results from \citet{Ricotti16}. The NSM runs have higher fractions of cold ($< 10^{3} K$) gas mass at later times, leading to this increase in Pop II star formation.

\begin{table}
	\centering
	\begin{tabular}{lcccc} 
		\hline
		 & \# Pop III & M$_{\mathrm{Tot, P2}} [\Ms]$ & $f_{\mathrm{r}}$ & $f_{\mathrm{r, high}}$ \\
		\hline
		Fiducial & 2 & $10^{4.87}$ & 80\% & 64\% \\
		D & 2 & $10^{5.27}$ & 80\% & 14\% \\
		E & 1 & $10^{4.80}$ & 74\% & 72\% \\
		Original & 2 & $10^{4.52}$ & -- & -- \\
		\hline
	\end{tabular}
	\caption[Information about the stellar populations within the explosion energy variation runs.]{Information about the stellar populations within the explosion energy variation runs. For each run, we list the number of Pop III stars that form, the total Pop II stellar mass at the final redshift, the mass fraction of Pop II stars that are r-process enhanced ($> 10^{-6} Z_{\mathrm{r}}$), and the mass fraction of Pop II stars that are highly r-process enhanced ($> 10^{-1} Z_{\mathrm{r}}$).}
	\label{tab:eng_stellar}
\end{table}

\begin{figure} 
	\includegraphics[width=\columnwidth]{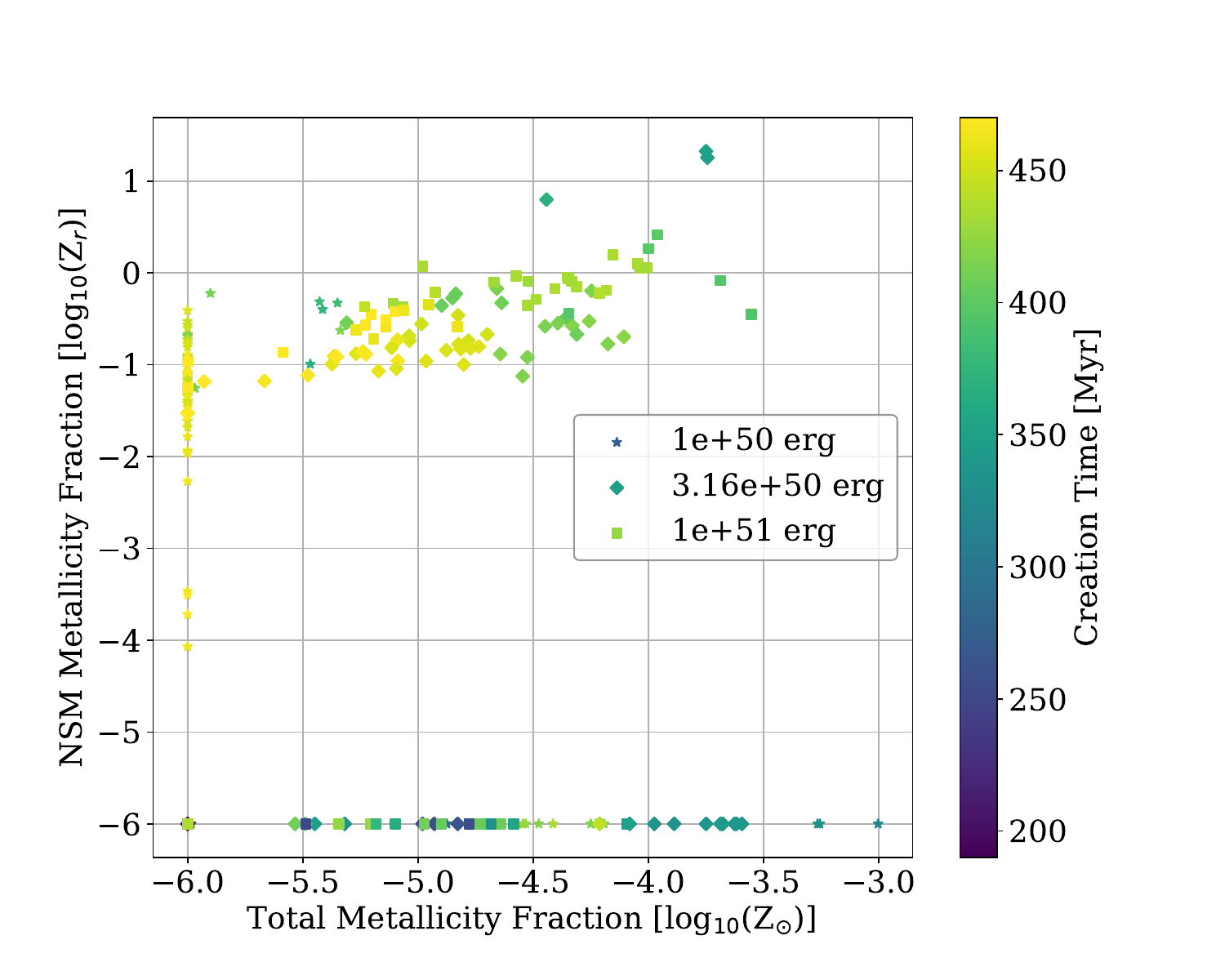}
	\caption[R-process metallicity fraction versus the total metallicity fraction for Pop II stars in the explosion energy variation runs.]{The r-process metallicity fraction versus the total metallicity fraction of Pop II stars at the final redshift. The different marker types are distinguished in the legend. The points are colored by their creation time. Note that the NSM occurs in these runs at $\sim 286$ Myr. We have also artificially raised the metallicity fraction for stars that have metallicity fractions below a value of $10^{-6}$ for clarity.}
	\label{fig:eng_NSM_p3}
\end{figure}

As the main halo assembles, and the neighboring halo falls in with its own pool of Pop II stars, the final halo ends with a mix of r-process enhanced and non-enhanced Pop II stars. This can be clearly seen in Figure \ref{fig:eng_NSM_p3}, where we are showing the NSM r-process metallicity fraction versus the total metallicity fraction, defined as the Pop III plus the Pop II metallicity fraction, of Pop II stars at $z = 10$. There is a collection of Pop II stars that are enriched with r-process material, and a separate group, at very low r-process metallicity fractions that formed either before the NSM occurred, or outside the enrichment zone. As Pop II stars form at later times, their r-process metallicity fractions decrease slightly, along with their total metallicity fraction. There is also a group of Pop II stars that form at very late times with high r-process metallicity fractions but low total metallicity fractions. 
 
\begin{figure} 
	\includegraphics[width=\columnwidth]{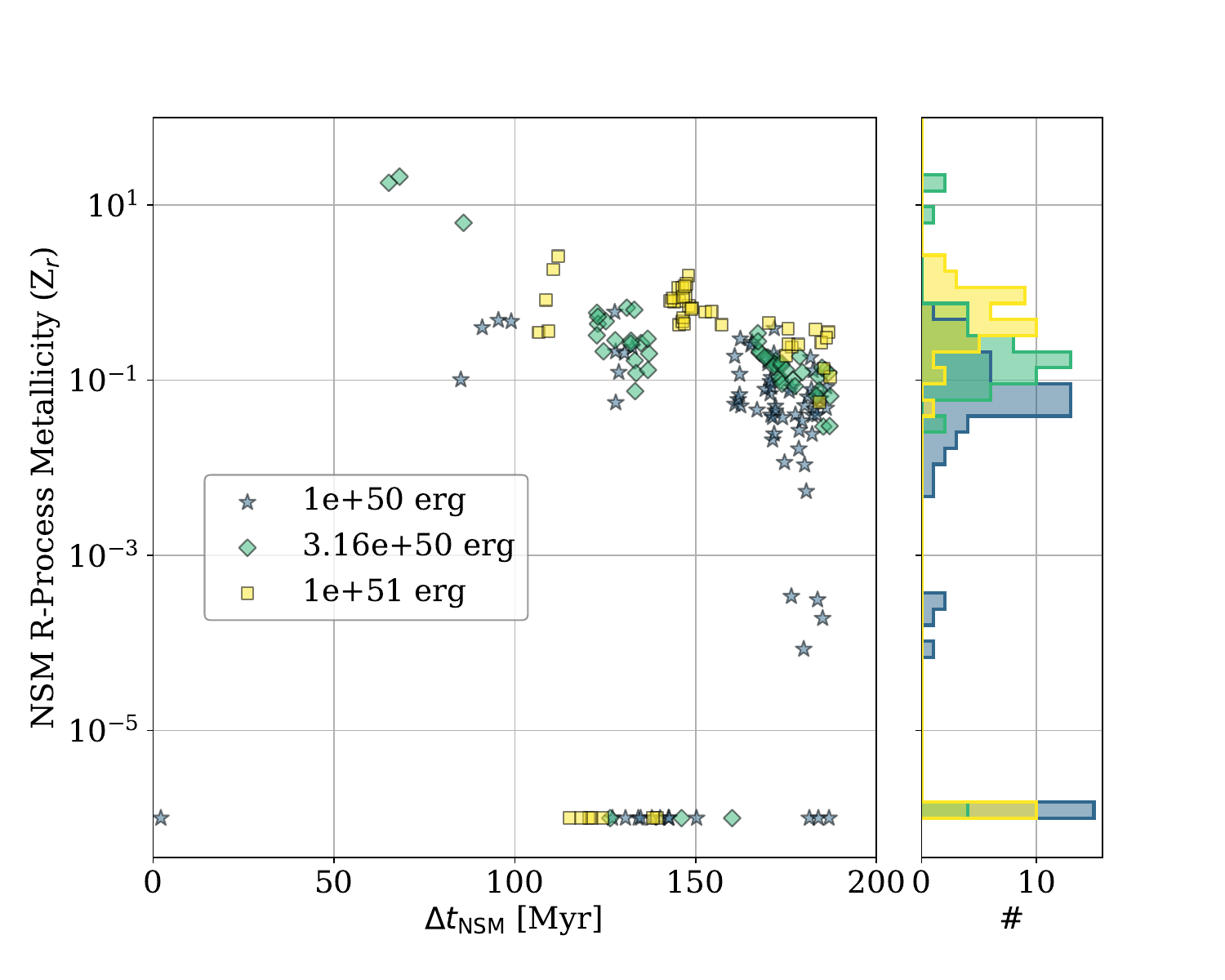}
	\caption[R-process metallicity as a function of $\Delta t_{\mathrm{NSM}}$ for the energy variation runs.]{The r-process metallicity for Pop II stars that formed in the halo as a function of when they formed relative to the NSM. The blue stars indicate Pop II stars formed in run D, the green diamonds from run Fiducial, and the yellow squares from run E. We have artificially raised the metallicity fraction for stars that have metallicity fractions below a value of $10^{-6}$ for clarity.}
	\label{fig:eng_NSM_deltat}
\end{figure}

Figure \ref{fig:eng_NSM_deltat} shows the r-process metallicity of the Pop II stars that formed within the halo as a function of their creation time relative to the NSM event. We define stars to be ``highly r-process enhanced'' if they have a metallicity of $> 10^{-1} Z_{\mathrm{r}}$ and to be ``r-process enhanced'' if they have a metallicity of $> 10^{-6} Z_{\mathrm{r}}$. At $z = 10$, we find that runs D, Fiducial and E have Pop II mass fractions of 80\%, 80\% and 74\% being r-process enhanced, and Pop II mass fractions of 14\%, 63\%, and 72\% being highly r-process enhanced, respectively. The lowest explosion energy results in a lower percentage of highly r-process enhanced stars, but a similar percentage of r-process enhanced stars compared to the higher explosion energy runs. As the energy increases, almost all Pop II stars that are enriched by the NSM are highly enhanced. This is due to the large amount of Pop II stars that form at late times near the center of the halo, where the majority of the r-process enhanced gas is contained. Figure \ref{fig:eng_NSM_deltat} also indicates that the Pop II stars that form closer to the NSM tend to have higher r-process metallicity fractions than if they were to form later on. We also see that the higher explosion energies produce Pop II stars with higher r-process metallicities, but less Pop II stars overall.

\subsection{Delay time variation runs} \label{sec:time_vary}

Runs A, B and Fiducial vary the delay time of the NSM by 10, 30 and 100 Myr, respectively. The NSM occurs in runs A, B and Fiducial at 196 Myr ($z = 18.7$), 216 Myr ($z = 17.6$) and 286 Myr ($z = 14.4$), respectively. Similar to the energy variation runs, the dynamics and visual appearance of the halos remains similar to the original run. Run A ends with a total Pop II stellar mass of $10^{5.25} \Ms$, with the mean and median creation time of Pop II stars being 455 Myr ($z = 10.3$) and 465 Myr ($z = 10.1$). A total of 6 Pop III stars form within the main halo progenitor of run A, with three forming at 474 Myr ($z = 10.0$) with masses of 43.3 \Ms{}, 27.7\Ms{}, and 12.5 \Ms{}, one at 369 Myr ($z = 12.0$) with a mass of 56.9 \Ms, one at 358 Myr ($z = 12.3$) with a mass of 20.4 \Ms{}, and the first forming at 151 Myr ($z = 22.6$) with a mass of 26.5 \Ms. All halo progenitors experience 29 Pop III SNe. By $z = 10$, there are still 21 Pop III stars living within the halo. Run B ends with a total Pop II stellar mass of $10^{5.8} \Ms$, with the mean and median mass weighted creation time of Pop II stars being 455 Myr ($z = 10.3$) and 458 Myr ($z = 10.2$). Three Pop III stars form within the main halo progenitor of run B, at 473 Myr ($z = 10.0$), 343 Myr ($z = 12.6$), and 151 Myr ($z = 22.6$) with masses of 17.9 \Ms{}, 152.7 \Ms{}, and 26.5 \Ms{}, respectively. All halo progenitors experience 32 Pop III SNe, and by $z = 10$, there are still 24 Pop III stars still living within the halo. We refer the reader to the top of the previous section for a reminder of these values for the Fiducial run. 

\begin{figure} 
	\includegraphics[width=\columnwidth]{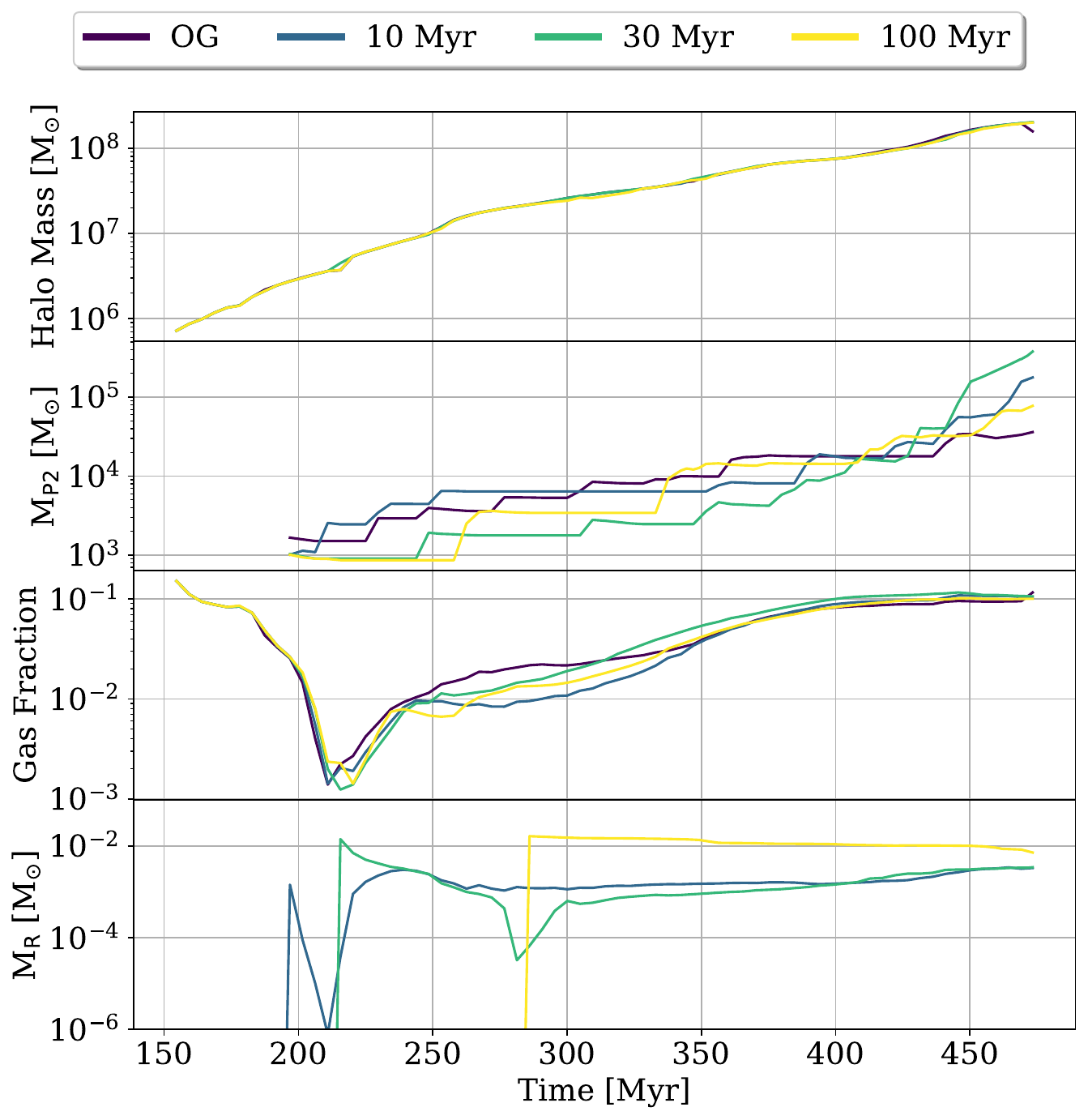}
	\caption[The halo mass, Pop II stellar mass, gas fraction, and r-process mass as a function of time for each delay time variation run.]{The evolution of the main halo in each delay time variation run and the original run. The top panel shows the halo mass, second from the top shows the Pop II stellar mass, third from the top shows the average gas fraction, and the bottom shows the r-process mass. The NSM occurs at 196 Myr, 216 Myr, and 286 Myr for runs A, B, and Fiducial, respectively,}
	\label{fig:time_evol}
\end{figure}

Figure \ref{fig:time_evol} shows the halo mass, the Pop II stellar mass, the gas fraction and the r-process mass as a function of time. The halo mass for these runs, like the explosion energy variation runs, are almost identical to each other, with the halos growing to $10^{8.3}$ \Ms{} for runs A and B. The Pop II stellar mass steadily increases as time increases, with run B ending with the highest Pop II stellar mass, followed by run A, then the Fiducial run, and finally the original run. The gas fraction in each run also follows a similar trend, with the first dip in gas fraction occurring due to the formation of the first stars when the halo was quite small, at $10^{6.6} \Ms$ for the original run and run A, and $10^{6.7} \Ms$ for runs B and Fiducial. The gas fraction steadily recovers, with the 30 Myr run overtaking the others with a higher gas fraction by 314 Myr. The bottom panel shows the r-process mass within the halo as a function of time, and the different runs have very different structures. The NSM in run A, when the delay time is 10 Myr, occurs very early when the halo was very small at a mass of $10^{6.4} \Ms$. The NSM ejects the r-process material out of the halo right away, which is why the initial r-process mass is smaller than the amount that is actually injected. In run B, when the delay time is 30 Myr, the r-process mass is mostly contained within the halo when the mass was $10^{6.7} \Ms$ at the time of the NSM but quickly starts to spread outward, past the bounds of the virial radius. A Pop III SN that occurs just on the boundary of the halo then carries this material even further outside the halo, resulting in the decrease in the r-process mass around 270 Myr. As merging takes place, some of this material is recovered, but an r-process ``tail'' remains outside of the halo. 

\begin{figure*} 
	\includegraphics[width=\textwidth]{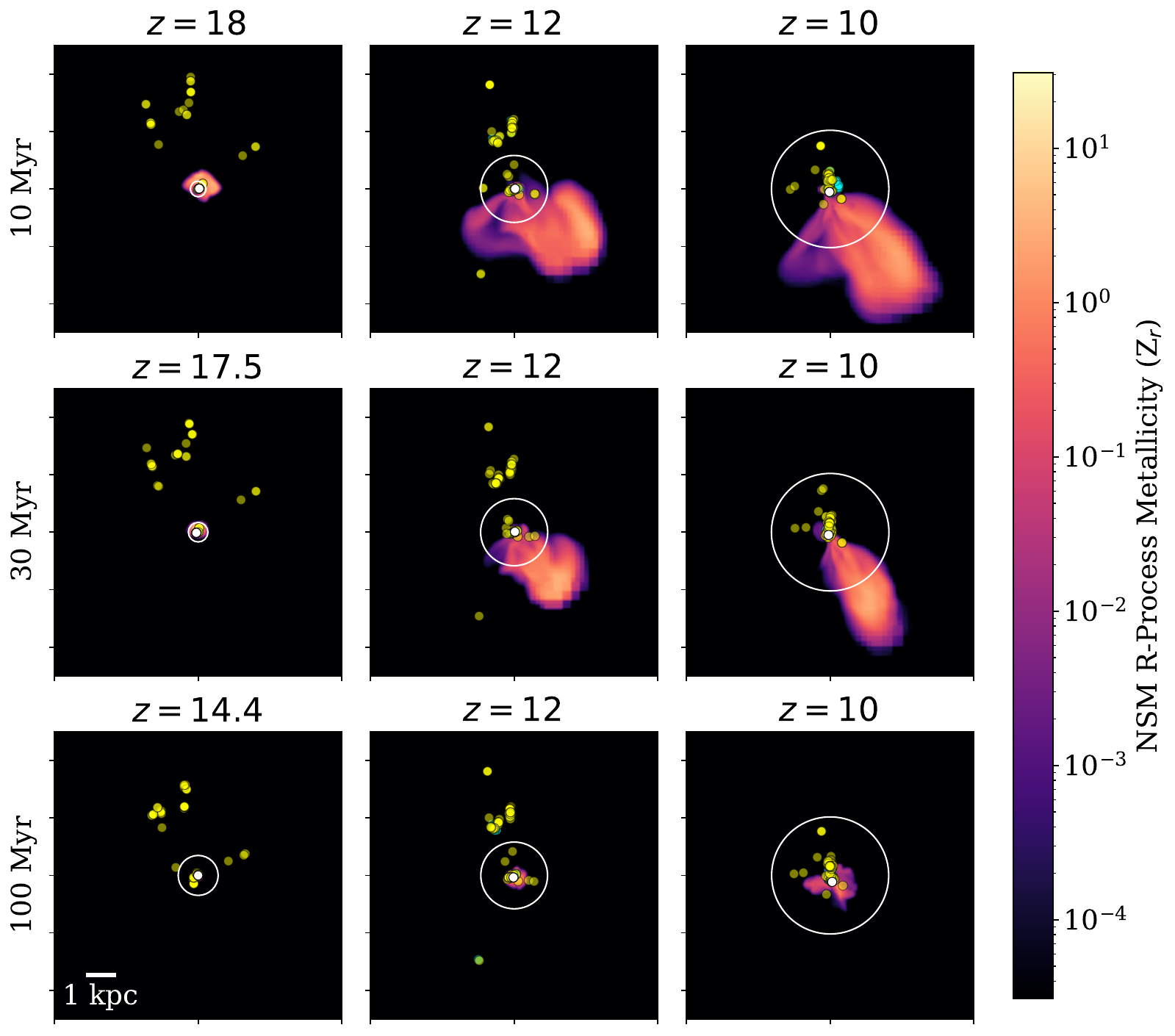}
	\caption[Projections of the r-process metallicity field for each delay time variation run through time.]{The r-process metallicity evolution for each delay time variation run. The first column shows the timestep at $z = 18, 17.5$ and $14.4$, when the NSM event takes place, the middle column shows $z = 12$, and the right column shows the final timestep of $z = 10$. The top row shows run A, the middle shows run B, and the bottom shows the Fiducial run. The white circle indicates the virial radius of the halo. The yellow dots show Pop III stars, and the aqua dots show Pop II stars. The white star indicates the chosen Pop III star that produced the NSM event. Note that the star particles are covering the NSM explosion in the Fiducial run at the first timestep.}
	\label{fig:time_met4_evol}
\end{figure*}

The evolution of the r-process metallicity field can be seen in Figure \ref{fig:time_met4_evol}. The top row represents run A, the middle run B, and the bottom the Fiducial run. The first column shows the timestep when the NSM takes place, which occurs at different times for these runs, based on their delay time. The middle column shows $z = 12$, and the right column shows the final redshift at $z = 10$. When the NSM occurs in run A, because the halo is so small at this early time, the r-process material is not contained within the halo, but extends outside of the halo. By $z = 12$, there are a number of Pop III stars that go SN and Pop II stars that form just outside of the halo, all of which pushes the gas outside of the halo even further through external feedback. As time progresses, some of the r-process material is recovered, but a large tail, or wing like structure remains outside of the halo. Run B follows a similar story as run A, with a similar structure appearing at $z = 10$. The total r-process mass is similar for these two runs by the end of the simulation. These short delay times appear very different as compared to the longer delay time, where the r-process material is always fully contained within the halo.

\begin{figure} 
	\includegraphics[width=\columnwidth]{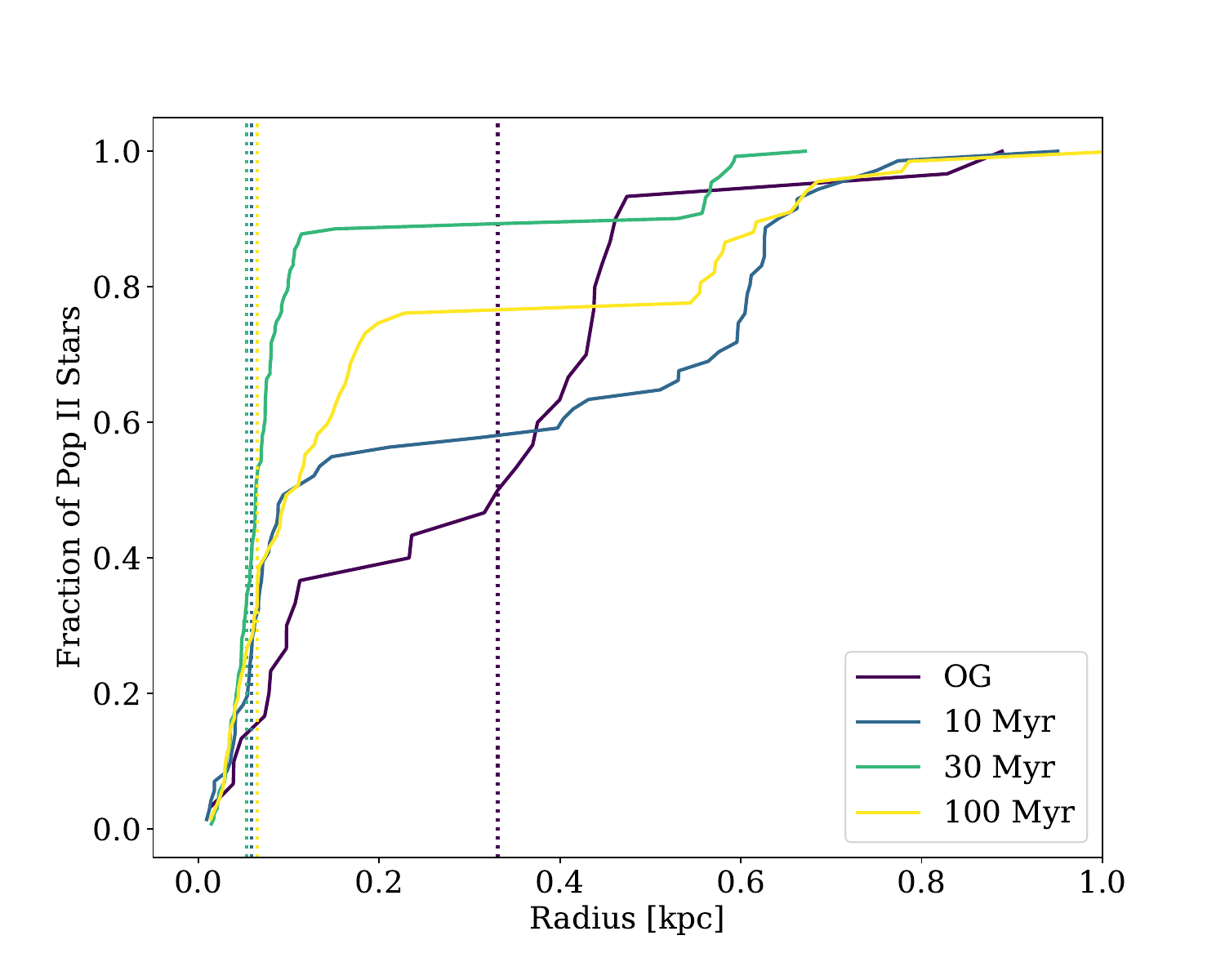}
	\caption[Radial distribution of Pop II stars in the delay time variation runs.]{The fraction of Pop II stars that are within some radius at the $z = 10$ for runs A, B and Fiducial, and the original run. The dashed lines indicate the half-mass radius, which is 331.2 pc, 59.2 pc, 53.3 pc, and 65.3 pc for the original run, run A, run B, and run Fiducial respectively. Note that the virial radius at this time is 1.6, 1.72, 1.72, and 1.73 kpc for the original run, run A, run B, and run Fiducial, respectively.}
	\label{fig:time_stellar_dist}
\end{figure}

Similar to the energy variation runs, the time delay runs lead to significantly higher Pop II stellar masses as compared to the original run. Runs A, B, and Fiducial end with Pop II stellar masses of $10^{5.25}\Ms$, $10^{5.58} \Ms$, and $10^{4.87}$ \Ms. Run A also produces six Pop III stars within the main halo progenitor, which is the highest number of Pop III stars formed within the main halo of all the runs. Table \ref{tab:time_stellar} presents this information in the first two columns. Figure \ref{fig:time_stellar_dist} shows the fraction of Pop II stars that exist within a certain radius for each run. The half-mass radius is indicated as the dotted line for each run and is 59.2 pc, 53.3 pc, and 65.3 pc for runs A, B, and Fiducial. There is a large cluster of Pop II stars that gathers near the center of the halo for each of the NSM runs, resulting in smaller half-mass radii. As with the energy variation runs, these half-mass radii are similar to those that have been found previously in simulations of UFD galaxies \citep{Ricotti16}. This larger group of Pop II stars originates from the NSM runs having higher cold gas mass fractions at late times as compared to the original runs. 

\begin{table}
	\centering
	\begin{tabular}{lcccc} 
		\hline
		 & \# Pop III & M$_{\mathrm{Tot, P2}} [\Ms]$ & $f_{\mathrm{r}}$ & $f_{\mathrm{r, high}}$ \\
		\hline
		Fiducial & 2 & $10^{4.87}$ & 80\% & 64\% \\
		A & 6 & $10^{5.25}$ & 70\% & 5\% \\
		B & 3 & $10^{5.58}$ & 74\% & 0.2\% \\
		Original & 2 & $10^{4.52}$ & -- & -- \\
		\hline
	\end{tabular}
	\caption[Information about the stellar populations within the delay time variation runs.]{Information about the stellar populations within the delay time variation runs. For each run, we list the number of Pop III stars that form, the total Pop II stellar mass at the final redshift, the mass fraction of Pop II stars that are r-process enhanced ($> 10^{-6} Z_{\mathrm{r}}$), and the mass fraction of Pop II stars that are highly r-process enhanced ($> 10^{-1} Z_{\mathrm{r}}$).}
	\label{tab:time_stellar}
\end{table}

\begin{figure} 
	\includegraphics[width=\columnwidth]{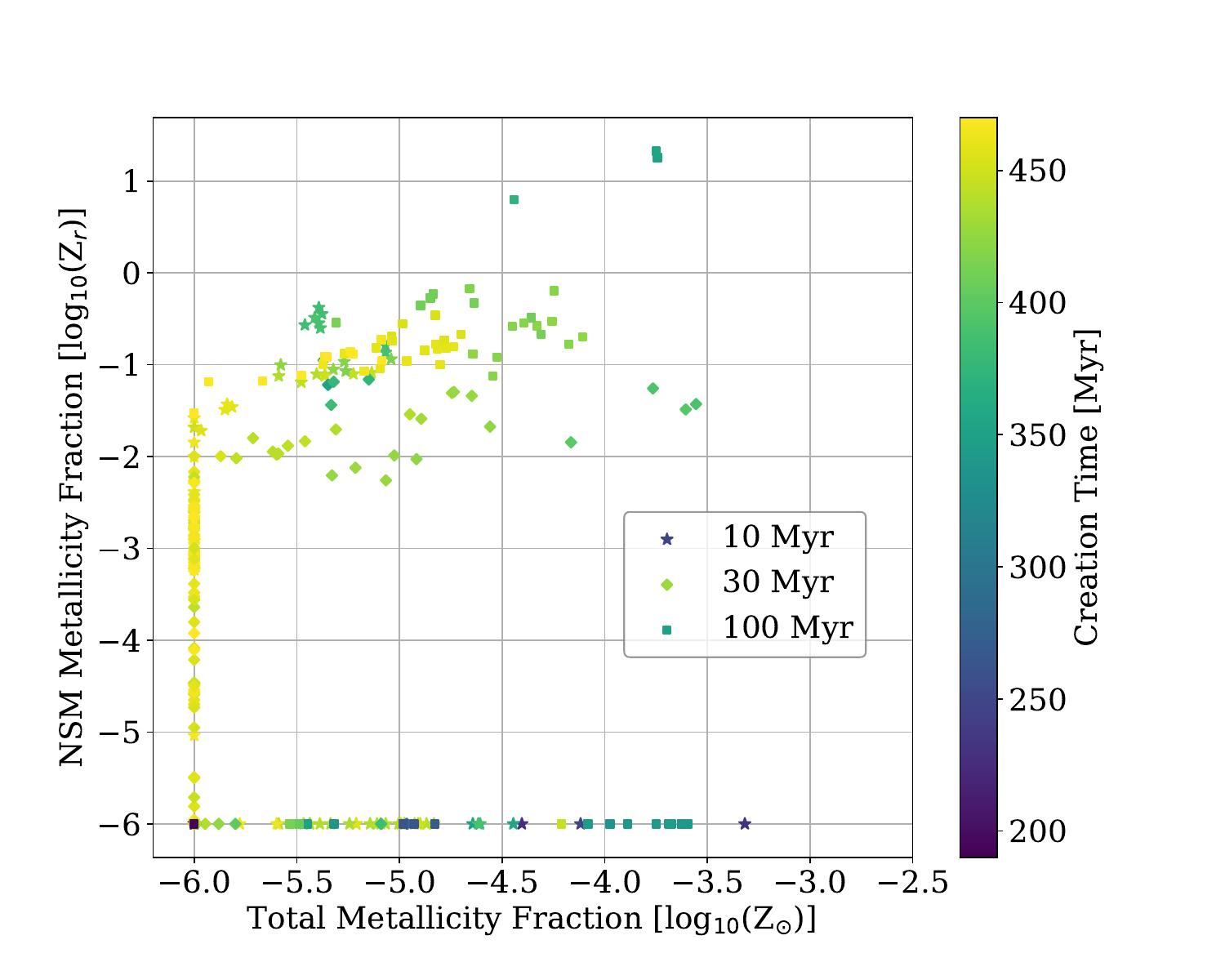}
	\caption[R-process metallicity fraction versus total metallicity fraction of Pop II stars in the delay time variation runs.]{The r-process metallicity fraction versus the total metallicity fraction of Pop II stars at the final redshift. The different marker types are distinguished in the legend. The points are colored by their creation time. Note that the NSM occurs in runs A, B, and Fiducial at 197 Myr, 216 Myr, and 286 Myr, respectively. We have also artificially raised the metallicity fraction for stars that have metallicity fractions below a value of $10^{-6}$.}
	\label{fig:time_NSM_p3}
\end{figure}

Figure \ref{fig:time_NSM_p3} shows the r-process metallicity fraction versus the total metallicity fraction for the Pop II stars at $z = 10$. Unlike the energy variation runs, there is slightly more scatter in the NSM metallicity fraction. There is a large portion of Pop II stars that form at low total metallicities but at relatively higher r-processed metallicities. This is again due to Pop II stars forming near the center of the halo where there is still a strong nucleus of r-process material.

\begin{figure} 
	\includegraphics[width=\columnwidth]{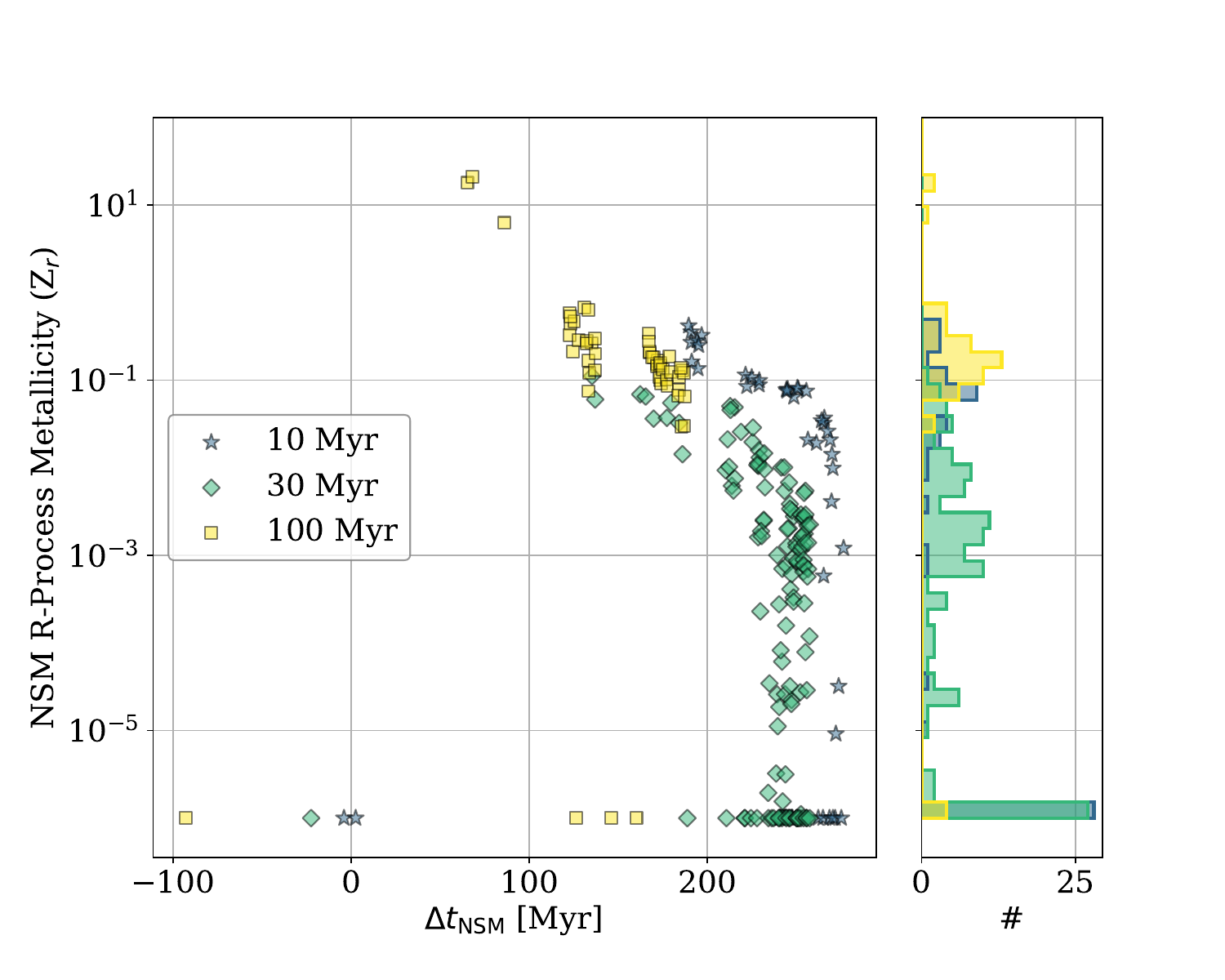}
	\caption[R-process metallicity fraction as a function of $\Delta t_{\mathrm{NSM}}$ for the delay time variation runs.]{The r-process metallicity for Pop II stars that formed in the halo as a function of when they formed relative to the NSM. The blue stars indicate Pop II stars formed in run A, the green diamonds from run B, and the yellow squares from run Fiducial. We have artificially raised the metallicity fraction for stars that have metallicity fractions below a value of $10^{-6}$ for clarity.}
	\label{fig:time_NSM_deltat}
\end{figure}

Figure \ref{fig:time_NSM_deltat} shows the r-process metallicity for Pop II stars as a function of their creation time relative to when the NSM occurred. In this plot, there are obvious groups of stars that come from the different runs. Run A, with the shortest delay time, forms Pop II stars at later times compared to when the NSM occurs, and at both high and low r-process metallicities. Run B, with the 30 Myr delay time, forms the highest amount of Pop II stars, but with relatively lower r-process metallicities overall. The Fiducial run forms Pop II stars the soonest after the NSM, which occurs simply because the delay time is longer, and with higher r-process metallicities. In this run, there hasn't been enough time for the material to escape the halo before Pop II Stars form, therefore, higher r-process metallicities are expected. We find that run A results in a 70\% and 5\% of the Pop II mass fraction being r-process enhanced and highly r-process enhanced, run B resulting in 74\% and 0.02\%, and run Fiducial resulting in 80\% and 64\% of the Pop II stars being r-process enhanced and highly r-process enhanced, respectively. This indicates that the NSM is in general enhancing a similar percentage of Pop II stars with r-process material, but the shorter delay times result in a much smaller percentage of Pop II stars being highly r-process enhanced.

\section{Discussion} \label{sec:discussion}

For all runs, we see similar Pop III stellar masses as compared to some recent studies. \citet{Riaz22} studied the Pop III contribution to high redshift galaxies for a range of halo masses. They find that a halo with a mass of $10^7$ \Ms{} has a Pop III mass fraction of $\sim 0.1$ at $z \approx 12$. For our runs, at a similar time and halo mass, we also see a Pop III mass fraction of $\sim 0.1$. By $z = 10$, we see $\sim 10^{4}$ \Ms{} of Pop II stars and $\sim 10^{3}$ \Ms{} for Pop III stars, as compared to $10^3 - 10^5$ \Ms{} of Pop II stars and  $10^1 - 10^2$ \Ms{} of Pop III stars found by \citet{Riaz22}. While we do see a slightly higher value of Pop III stellar masses as compared to them, our overall mass fractions are similar. \citet{Yajima23} studied the first galaxies and their Pop III contributions in order to compare with results from JWST. They also look at the Pop III mass fraction (see their Figure 5) as compared to the total stellar mass. They report a much broader range of values for different redshifts in general. For a stellar mass of $10^{4.5}$ \Ms, they find a Pop III mass fraction of $\sim 0.18$, while we see a mass fraction for that stellar mass around 0.03. However, we do see a similar trend, that as stellar mass increases, the Pop III mass fraction begins to decrease. At the end of our simulations when our halo has a stellar mass of $\sim 10^{5.5}$ \Ms, we find a Pop III mass fraction of $\sim 0.006$, which is similar to \citet{Yajima23}. Based on these results, we agree with their conclusion that for halos with stellar masses of $\sim 10^{7}$ \Ms, the Pop III mass fraction is small, at $< 0.01$.

This work represents some of the first simulations on the chemical and dynamical impacts of a NSM, with various parameters, in a cosmological setting, and thus there are some difficulties that come with this kind of study. The first is that the small sample size prevents us from saying anything conclusive about the NSM parameters in general. While we see some trends across the data, the inherent randomness across simulations leads to different amounts of star formation at different times and locations. On top of that, because we randomly sample the Pop III IMF, if two simulations are looking very similar and both go on to form a Pop III star in a similar location, one can be sampled at a low mass and the other at a very high mass, and affect the surrounding region in completely different ways. What may seem like a small difference in the Pop III stellar mass can go on to have impactful consequences on future star formation. This means that it can be difficult to pick apart why certain stars form where and at what times. We stress that this is merely a proof-of-concept study to show that we can study the impacts of NSMs in a cosmological setting to present meaningful results that can enhance our understanding of r-process enrichment in the early universe. We plan to follow up this paper with a study that separates the source of differences between the simulations, and are outside the scope of this paper.

The NSM model we implement here is relatively simple in order to understand the impacts of only a small amount of parameters. We manually choose which Pop III star to become the Pop III binary system, and thus the NSM, and follow only a general r-process metal field to $z = 10$. The simple model allows us to focus on just the r-process enrichment of the halo in order to better understand how this simple model behaves and how the r-process field affects the second generation of stars. But because it is relatively simple, we lack the full details of NSMs in the universe. Some details that could be included would be to automatically form a NSM given some NSM rate, to include asymmetric NSMs and kilonova effects, and to follow individual metal species. These are all next steps to what is a long endeavor in this field of simulating NSMs in a cosmological setting. 

Finally, if we want to get a better sample size of how NSMs affect the second generation of stars, many simulations would need to be run, or a single larger simulation would need to be executed, in which case the resolution that can be achieved in zoom-in simulations are lost. Simulations are in general not cheap to run, but to better understand the impacts of these parameters on star formation, a larger sample size is needed. Implementing an automatic NSM model within cosmological simulations that assigns various explosion energies and delay times would better help us constrain the impacts of these events on metal-enriched star formation in the early universe.
	
\section{Conclusions} \label{sec:conclusion}

We have run a suite of simulations studying the impact of Pop III remnants, in the form of NSMs, on the second generation of star formation in the universe. We ran one simulation without a NSM, and five simulations varying the explosion energy and the delay time of the NSM in order to see how these variations affect the r-process abundances of the next generation of stars. In our case study, we find that:
\begin{itemize}
	\item NSMs originating from Pop III stars do lead to a substantial fraction of highly r-process enhanced Pop II stars.
	\item When the explosion energy is high, 73\% of stars are highly r-process enhanced, versus when the explosion energy is lower, only 14\% are highly r-process enhanced. 
	\item The lower explosion energy leads to a higher fraction of Pop II stars being r-process enhanced overall, but at lower metallicities in general. The higher explosion energy leads to a smaller fraction of r-process enhanced stars, but almost all that are enhanced are highly enhanced. 
	\item The longer time delays lead to higher mass fractions of highly r-process enhanced stars, due to the NSM occurring closer to the period of Pop II star formation. 
	\item The shorter time delays still lead to a high mass fraction of r-process enhanced stars, but a much smaller fraction of highly r-process enhanced stars.
	\item The half-mass radius significantly decreases in each NSM run due to the large number of Pop II stars that form and cluster near the center of the halo at the end of the simulation. 
	\item Star formation and feedback that occur in and around the halo heavily govern the dynamics and details of star formation in these runs. A larger sample size of galaxies with NSMs is needed in order to more fully understand the impact of a NSM on the second generation of stars. 
\end{itemize}

Given that we have shown that this simple NSM works well to study how a NSM might affect star formation in the early universe, our next steps are to run the model on a simulation of an UFD progenitor halo in order to attempt to recreate Ret II. These simulations are ongoing and will be an extension of this work. After that, improving the NSM model and following individual metal species, as discussed in Section \ref{sec:discussion} would greatly enhance the detail of this work. This work shows that NSMs can lead to significant amounts of r-process enrichment in Pop II stars, and will aid our understanding of the source of r-process materials in the universe and the nature of Pop III stars.

\section*{Acknowledgements}

DS is supported by the NASA FINESST fellowship award 80NSSC20K1540. The figures in this paper were constructed with the plotting library \textsc{matplotlib} \citep{matplotlib}.  JHW is supported by NSF grants OAC-1835213 and AST-2108020 and NASA grants 80NSSC20K0520 and 80NSSC21K1053.

\section*{Data Availability}

The data presented in this work is available upon reasonable request to the corresponding author.


\bibliographystyle{mnras}
\typeout{}
\bibliography{drenniks} 








\bsp	
\label{lastpage}
\end{document}